\documentclass[aps,prb,twocolumn,showpacs]{revtex4-1}
\usepackage{amsmath,amssymb,amsfonts}
\usepackage{graphicx}
\usepackage{dcolumn}
\usepackage{bm}
\usepackage[colorlinks=true,linkcolor=blue]{hyperref}%

\begin{document}

\title{Critical surface band gap of repulsive Casimir interaction between three dimensional topological insulators at finite temperature}

\author{Liang Chen}
\author{Shaolong Wan}
\email[]{slwan@ustc.edu.cn}
\affiliation{Institute for Theoretical Physics and Department of Modern Physics\\
University of Science and Technology of China, Hefei, 230026, People's Republic of China}

\date{\today}

\begin{abstract}
We generalize the calculation of Casimir interaction between topological insulators with opposite topological magnetoelectric polarizabilities and finite surface band gaps to finite Temperature cases. We find that finite temperature quantitatively depress the repulsive peak and enlarge the critical surface gap $m_c$ for repulsive Casimir force. However the universal property $m_c a \sim 1/2$ is still valid for various oscillation strength, temperature region and topological magnetoelectric
polarizabilities.
\end{abstract}


\maketitle

\section{\label{sec1}Introduction}

Exploration for exotic physical properties about topological protected quantum states is an important theme of current condensed matter physics. The recently discovered topological insulator(TI)
\cite{QiphysToday2010,Hasanrmp2010,Moorenature2010}
is such a quantum state. The three dimensional topological insulator has a bulk gap like an normal insulator, however the surface state of this material is gapless
\cite{Konigscience2007,Hsiehnature2008,HJZhangnature2009,Hsiehnature2009,Hsiehprl2009,YXianature2009,YLChenscience2009},
and such a gapless spectrum together with odd Dirac cones on TI surface are topological protected by the time-reversal symmetry
\cite{Fuprl2007,Fuprb2007,Mooreprb2007,Qiprb2008}.
There are many interesting phenomenas(predictions) related to this novel material, such as the topological magnetoelectric effect
\cite{Qiprb2008},
electric charge induced magnetic monopole
\cite{Qiprb2008,Qiscience2009},
optical Kerr and Faraday rotation
\cite{Tseprl2010,Josephprl2010,Sushkovprb2010},
surface $1/2$ quantum Hall effect
\cite{Qiprb2008,Chuprb2011}, \textit{et.al}.

Casimir effect is a quantum effect arising from zero-point energy fluctuation of vacuum, the seminal work of H. B. G. Casimir
\cite{Casimir1948}
found that two parallel uncharged metallic planes will emergence an attractive force. Before investigated in TI systems, casimir force has been proposed to be repulsive for some special conditions. For instance, it is proposed that Casimir force is repulsive if special geometry has been considered
\cite{Levinprl2010},
it is also reported that Casimir interaction between metamaterials maybe repulsive
\cite{Zhaoprl2009,Zhaoprb2011},
experimental evidence
\cite{Mundaynature2009}
shown that high-refractive liquid
\cite{Zwolpra2010}
between dielectrics will induce repulsive Casimir force.

Recently, A. G. Grushin and A. Cortijo proposed
\cite{GrushinPRL,Grushinprb2011}
that Casimir interaction between TIs with opposite topological magnetoelectric polarizability is repulsive while the distance between TIs tends to zero. Their analyzation is based on the topological quantum field description of TI
\cite{Qiprb2008},
$S_{topo} = \alpha/(4\pi)^2 \int{d^3}x{d}t\theta\bm{E}\cdot\bm{B}$, where $\alpha$ is the fine structure constant, $\theta=(2n+1)\pi$ is the topological magnetoelectric polarizability, $\bm{E}$ and $\bm{B}$ are electric and magnetic field respectively. Such a topological quantum field description is exact, however, the repulsive Casimir force will be suppressed by conducting surface fermions. In order to deduce the influence of surface fermions, one need to open a surface band gap by adding a magnetic coating on TI. We analyzed the Casimir interaction between TIs for finite surface band gap at zero temperature
\cite{LChenprb2011}.
We found that a critical surface band gap $m_c$ is essential for repulsive Casimir interaction, and such a surface band gap can be estimated by $m_c a\sim1/2$, where $a$ is the distance between TIs.

For practical measurement, the effect of temperature is always need to be considered. In this paper, we calculated the Casimir interaction between TIs with opposite topological magnetoelectric polarizability at finite temperature, we found that the general relation $m_c a\sim1/2$ is still valid.

This paper is organized as follows: In Sec.\ref{sec2}, we derive an effective action of surface electromagnetic field by integration out the contribution of surface fermions with finite surface band gap. From the effective action, we deduce the Maxwell equations with boundary corrections and Fersnel coefficient matrix in Sec.\ref{sec3}. In Sec.\ref{sec4}, we calculate the Casimir energy between TIs by Lifshtz formula, then we present the scope of repulsive Casimir force for different temperatures, surface band gap and topological magnetoelectric polarizabilities. Conclusions are given in Sec.\ref{sec5}.\\

\section{\label{sec2}Effective Action at finite temperature}

Let us formulate the model, in the vacuum and bulk of TIs, the action of electromagnetic field can be written as:
\begin{equation}
\label{eq1}
S = -\frac{1}{8\pi}\int dt d^3x \left(\varepsilon\bm{E}^2 -
\frac{1}{\mu}\bm{B}^2\right), 
\end{equation}
where $\bm{E}$ and $\bm{B}$ are electric and magnetic field, $\varepsilon$ and $\mu$ are permittivity and permeability of TI in the bulk and equal to 1 in the vacuum.

The topological nontrivial term $\alpha/(4\pi)^2 \int{d^3}x{d}t\theta\bm{E}\cdot\bm{B}$ can be modeled by massive surface Dirac fermions:
\begin{equation}
\label{eq2}
S_D = \int {d^3}x \bar{\psi}\left[ i\gamma^{a} (\partial_{a} +
ieA_{a} -m )\right]\psi, 
\end{equation}
where $a=0,x,y$; $\gamma^{0}=\sigma^{z}$, $\gamma^{x}=iv_F\sigma^{y}$, $\gamma^{y}=-iv_F\sigma^{x}$ and $\bar{\psi}=\psi^{\dagger}\gamma^{0}$. $\sigma^{x,y,z}$ are the
three Pauli matrices of the spin, and $v_F$ is the Fermi velocity of the surface fermion, it takes different values for different materials
\cite{YLChenscience2009,YXianature2009},
for example, $v_F=1.3\times10^{-3}$ for Bi$_2$Te$_3$, $v_F=1.7\times10^{-3}$ for Bi$_2$Se$_3$, in this paper, we take $v_F=1.0\times10^{-3}$ for numerical calculation. $A_a$ are the first three components of the electromagnetic potential. $m$ is the surface band gap opened by magnetic coating, $m=\pm|m|$ corresponding to $\theta=\pm\pi$. The generalization to $\theta=(2n+1)\pi$ is straightforward by introducing multi-fermions on TI surfaces. For analytical derivation, we only consider the case $\theta=\pm\pi$, the general case will be considered in Sec.\ref{sec4}.

\begin{figure}
\includegraphics[width=0.45\textwidth]{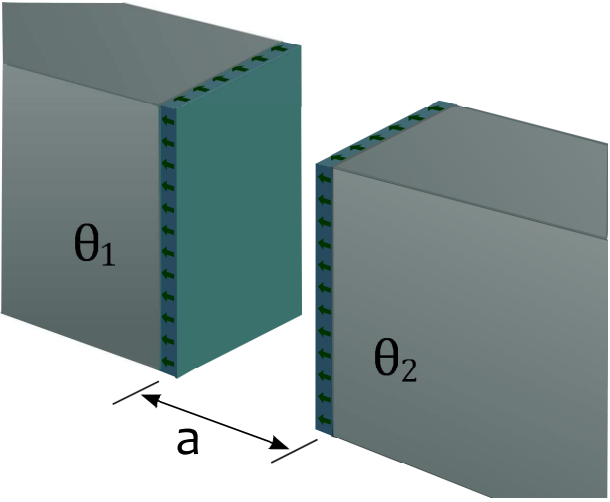}%
\caption{Schematic illustration of Casimir interaction between TIs
with opposite topological magnetoelectric polarizability
\(\theta_{1}=-\theta_{2}\). We assume the thickness of magnetic
coating is much smaller than the distance between TIs.\label{figure_1}}
\end{figure}

Formally, one can integrate out the contribution of surface fermion to get an effective action of electromagnetic field on TI surface,
$S_{eff}(A)=-i\ln\det\left[i\gamma^{a}(\partial_a+ieA_a)-m\right]$.
Up to one-loop approximation, the quadratic term of effective action $S_{eff}(A)$ can be written as:
\begin{equation}
\label{eq3}
S_{eff}(A)=\frac{1}{2}\int\frac{d\omega}{2\pi}\frac{d^2k}{(2\pi)^2}
A_{a}(\bm{k},\omega) \Pi^{ab}(\bm{k},\omega) A_{b}(\bm{k},\omega).
\end{equation}

For the detailed derivation of polarization operator tensor $\Pi^{ab}$ at finite temperature, we work in Matsubara imaginary time formalism:
\begin{equation}
\label{eq4}
i\Pi^{ab}(k) = \frac{e^2}{\beta}\sum_{m} \int
\frac{d^2p}{(2\pi)^2}\text{Tr}\left[\gamma^{a}G(k+p)\gamma^{b}
G(p)\right], 
\end{equation}
where $\beta=1/{k_{B}T}$ is the inverse of temperature and $k_B$ is the Boltzmann constant, $k=(i\omega_n, \bm{k})$, $p=(i\omega_m, \bm{p})$, $G(p)=i/(\gamma^{a}k_a+m)$ is the propagator of the surface fermion. $i\omega_n = 2n{\pi}i/\beta$ and $i\omega_m=(2m+1){\pi}i/\beta$ are the finite temperature frequency of electromagnetic field and surface fermion respectively.

The action of surface fermions are relativistic and satisfies Lorentz symmetry(if we set the Fermi velocity $v_F=1$), a similar action and corresponding polarization tensor have been considered in graphene system
\cite{Fialkovskyprb2011}
and 3-dimensional quantum electromagnetic dynamics
\cite{WLiprd2010},
the only difference here is that we have only one specie of Dirac fermion here, so that the topological parity odd term is preserved, after derivation, we find
the polarization tensor can be divided into three parts:
\begin{equation}
\label{eq5}
\Pi(k) = 2\Phi_{1} \Pi_{S1}(k) + 2v_F^2 \Phi_{2} \Pi_{S2}(k) - i
\phi \Pi_{AS}(k), 
\end{equation}
where $\Pi_{S1}(k)$ and $\Pi_{S2}(k)$ are parity even and $\Pi_{AS}(k)$ are parity odd, their exact forms are:
\begin{eqnarray}
\label{eq6}
\Pi_{S1}(k) &=& \left(
                \begin{array}{ccc}
                  k_x^2+k_y^2 & -i\omega_{n}k_x & -i\omega_{n}k_y \\
                  -i\omega_{n}k_x & -\omega_n^2 & 0 \\
                  -i\omega_{n}k_y & 0 & -\omega_n^2 \\
                \end{array}
              \right), \nonumber\\
\Pi_{S2}(k) &=& \left(
                \begin{array}{ccc}
                  0 & 0 & 0 \\
                  0 & -k_y^2 & k_xk_y \\
                  0 & k_xk_y & -k_x^2 \\
                \end{array}
              \right), \nonumber\\
\Pi_{AS}(k) &=& \left(
                  \begin{array}{ccc}
                    0 & -k_y & k_x \\
                    k_y & 0 & -i\omega_n \\
                    -k_x & i\omega_n & 0 \\
                  \end{array}
                \right), 
\end{eqnarray}
where $\Phi_1$, $\Phi_2$ and $\phi$ are three parameters, which can be derived from Eq.\ref{eq4} straightforward via Feynman parametrization and redefining the integration variable $\bm{l}=\bm{p}+x\bm{k}$:

\begin{eqnarray}
\label{eq7}
\Phi_1&=&-\frac{1}{\bm{k}^2}\frac{e^2}{\beta}\sum_m \int_0^1 {d}x \int \frac{d^2l}{(2\pi)^2} \nonumber\\
 & &\frac{\left(m^2+v_F^2\left(\bm{l}^2-x(1-x)\bm{k}^2\right)
-\omega_m\left(\omega_m +
\omega_n\right)\right)}{\left[\left(\omega_m+x\omega_n\right){}^2+v_F^2\bm{l}^2+\Delta\right]^2}, \nonumber\\
\Phi_2&=&\frac{e^2 v_F^2}{\beta}\sum_m \int_0^1{d}x \int
\frac{d^2l}{(2\pi)^2}
\frac{2x(1-x)v_F^2}{\left[\left(\omega_m+x\omega_n\right){}^2+v_F^2\bm{l}^2+\Delta
\right]^2}, \nonumber\\
\phi&=&\frac{e^2}{\beta}\sum_m \int_0^1{d}x \int
\frac{d^2l}{(2\pi)^2} \frac{2{v}_F^2
m}{\left[\left(\omega_m+x\omega_n\right){}^2+v_F^2\bm{l}^2+\Delta\right]^2},
\end{eqnarray}

where $\Delta  = m^2+x(1-x)\left(\omega_n^2+v_F^2\left(k_x^2+k_y^2\right)\right)$. One can carry out the integration over momentum and summation over frequency, and get the form of these parameters with only the integration over Feynman parameter $x$:

\begin{eqnarray}
\label{eq8}
\Phi_1&=&-\frac{\alpha{T}}{v_F^2\bm{k}^2}\int_0^1{d}x
\left\{\left[f^+\tanh(\lambda^+)-\log\Big(2\cosh\left(\lambda^+\right)\Big)\right]\right. \nonumber\\
&&\left.\text{\hspace{0.5cm}}+\left(\lambda^+\to\lambda^-,f^+\to
f^-\right)\right\}, \nonumber\\
\Phi_2&=&\frac{\alpha}{2} \int_0^1 {d}x \frac{
x(1-x)}{\sqrt{\Delta}}\left[\tanh(\lambda ^+)+\left(\lambda ^+\to
\lambda ^-\right)\right], \nonumber\\
\phi&=&\frac{m\alpha}{2}\int_0^1\frac{{d}x}{\sqrt{\Delta}}\left[\tanh(\lambda^+)+\left(\lambda^+\to
\lambda^-\right)\right], 
\end{eqnarray}

where $\alpha$ is the fine structure constant and

\begin{eqnarray}
\label{eq9}
\lambda^{\pm}&=&{\left(\sqrt{\Delta }\pm x i\omega _n\right)}/{(2T)}, \nonumber\\
f^{\pm}&=&\frac{2\Delta\mp(1-2 x)i\omega_n\sqrt{\Delta
}-2x(1-x)v_F^2\left(k_1^2+k_2^2\right)}{4 T\sqrt{\Delta}}. \nonumber\\
\end{eqnarray}

By using the series expansion $\tanh{t}=1+2\sum_{j=1}^{\infty}(-1)^{j} e^{-2t}$ and $\log(2\cosh{t})=t-\sum_{j=1}^{\infty} \frac{(-1)^j}{j}e^{-2t}$, one can rewrite these three parameters as

\begin{widetext}
\begin{eqnarray}
\label{eq10}
\Phi_1 &=& \alpha\int_0^1{d}x \frac{x(1-x)}{\sqrt{\Delta}}
-4\frac{\alpha{T}}{v_F^2\bm{k}^2} \int_0^1 \text{dx}
\sum_{j=1}^{\infty}(-1)^{j} e^{-j\sqrt{\Delta}/T}
\left[\left(\text{Re}{f^+}-\frac{1}{2j}\right)\cos(2j\pi{nx}) +
\text{Im}{f^+} \sin(2j\pi{nx})\right], \nonumber\\
\Phi_2 &=& \alpha\int_0^1{d}x
\frac{x(1-x)}{\sqrt{\Delta}}\left[1+2\sum _{j=1}^{\infty }
(-1)^je^{-j \sqrt{\Delta }/T}\cos(2j n\pi x)\right], \nonumber\\
\phi  &=& \alpha  m\int_0^1{d}x \frac{1}{\sqrt{\Delta
}}\left[1+2\sum _{j=1}^{\infty } (-1)^je^{-j \sqrt{\Delta
}/T}\cos(2j n\pi x)\right], 
\end{eqnarray}
\end{widetext}

where $\text{Re}f^{+}$ and $\text{Im}f^{+}$ take the real and imaginary part of $f^{+}$. It is easy to check that in the low temperature limit $(T\rightarrow0)$, these expressions coincide with the zero temperature results. By using Eq.\ref{eq3}, we get the effective Lagrangian of surface electromagnetic field:

\begin{eqnarray}
\label{eq11}
\mathcal{L}_S&=&-\frac{\phi}{8\pi}\epsilon_{abc}A^{a}\partial^{b}A^{c}\nonumber\\
&&+\frac{1}{4\pi}\left(\Phi_1 \sum_{j=x,y}{F}_{0j}{F}^{0j}+\Phi_2
v_F^2{F}_{\text{xy}}{F}^{\text{xy}}\right), 
\end{eqnarray}

where $F_{ab}=\partial_{a}A_{b}-\partial_{b}A_{a}$ is the electromagnetic field tensor.\\

Combine Eq.\ref{eq1} and Eq.\ref{eq11}, we obtain an action of the whole system:

\begin{eqnarray}
\label{eq12}
S &=& \int {d}t{d^3}x \left\{ -\frac{1}{8\pi}(\bm{E}^2-\bm{B}^2)\theta(z)\theta(a-z) \right. \nonumber\\
&&-\left.\frac{1}{8\pi}\left( \varepsilon\bm{E}^2-\frac{1}{\mu}\bm{B}^2\right)(\theta(-z)+\theta(z-a)) \right.\nonumber\\
&&-\left.\delta(z_i)\left[\frac{\phi_i}{8\pi}\epsilon_{abc}A^{a}\partial^{b}A^{c}
-\left(\frac{\Phi_1}{4\pi}\sum_{j=x,y}{F}_{0j}{F}^{0j}\right.\right.\right.\nonumber\\
&&+\left.\left.\left.v_F^2\frac{\Phi_2}{4\pi}
{F}_{\text{xy}}{F}^{\text{xy}}\right) \right]\right\},
\end{eqnarray}

where $a$ is the distance between TIs, and we have omitted the thickness of magnetic coating(see schematic illustration in Fig.\ref{figure_1}), $\theta(t)$ is the Heaviside unit step function, $i=1,2$ and $z_1 = 0$, $z_2=a$. $\phi_i$ is the value of $\phi$ for different surfaces, without lose of generality, we have assumed the absolute value of surface band gap equal to each other on the two surfaces so that different signs of surface band gap corresponding to different signs of the topological term
$\alpha\theta\bm{E}\cdot\bm{B}/(4\pi)^2$ in the topological quantum field description of TIs. We also note that the effect of finite temperature has been implicitly included in parameters $\Phi_1$, $\Phi_2$ and $\phi$.\\

\section{\label{sec3}Modified Maxwell equations and Casimir interaction}

The Euler-Lagrange equations of the action\ref{eq12} give the Maxwell equations of electromagnetic field with surface corrections:

\begin{eqnarray}
\label{eq13}
&&\frac{1}{4\pi}\triangledown\cdot\bm{D}=-\delta(z-z_i)\left(\frac{\phi_i}{4\pi}{B}_z-\frac{\Phi_1}{2\pi
}\triangledown\cdot\bm{E}\right), \nonumber\\
&& \frac{1}{4\pi}\left(\frac{\partial\bm{D}}{\partial
t}-(\triangledown \times \bm{H})\right)= \nonumber\\
&&\delta(z-z_i)\left(\frac{\phi_i}{4\pi}\tilde{\bm{E}}+\left(\frac{\Phi_1}{2\pi}\frac{\partial\bm{E}}{\partial
t}-\frac{\Phi_2 v_F^2}{2\pi} \triangledown\times\bm{H}\right)\right), \nonumber\\
&& \triangledown \cdot\bm{B} =0, \nonumber\\
&& \frac{\partial\bm{B}}{\partial t}+(\triangledown \times
\bm{E})=0, 
\end{eqnarray}

where $\bm{D}=\varepsilon\bm{E}$ and $\bm{H}=\bm{B}/\mu$ are electric displacement field and magnetizing field; $\tilde{E}_i=\epsilon_{ij}E_j$($i,j=x,y$). From these modified Maxwell equations, we get the discontinuous boundary conditions:

\begin{eqnarray}
\label{eq14}
{D}_z\left(z_i^+\right)-{D}_z\left(z_i^-\right)&=&
-{\phi_i}{B}_z+2{\Phi_1}\triangledown \cdot\bm{E}, \nonumber\\
{H}_x\left(z_i^+\right)-{H}_x\left(z_i^-\right)&=&{\phi_i}{E}_x-2\left({\Phi_1}{\partial_t{E}_y}
+{\Phi_2v_F^2}\partial_x{B}_z\right), \nonumber\\
{H}_y\left(z_i^+\right)-{H}_y\left(z_i^-\right)&=&{\phi_i}{E}_y+2\left({\Phi_1}{\partial_t{E}_x}
-{\Phi_2v_F^2}\partial_y{B}_z\right), \nonumber
\end{eqnarray}

where $z_i^{\pm}$ means $z_i\pm0$. The other three components $E_x$, $E_y$ and $B_z$ are continuous on the interface. From the discontinuous boundary conditions we find that a TE mode injection will induce both TE and TM mode reflection/refraction. The electromagnetic waves with injection TE mode in the vacuum can be written as:

\begin{eqnarray}
\label{eq15}
\bm{E} &=& (1+r_{ee})\omega(-k_y\bm{e}_x+k_x\bm{e}_y) +
r_{em}(-k_z\bm{k}-k^2\bm{e}_z), \nonumber\\
\bm{B} &=& (-k_z\bm{k}+k^2\bm{e}_z) + r_{ee}(k_z\bm{k}+k^2\bm{e}_z) \nonumber \\
&& + r_{em} \omega(-k_y\bm{e}_x+k_x\bm{e}_y), 
\end{eqnarray}

where $r_{ee}$ and $r_{em}$ are reflection coefficients of TE and TM mode respectively, the refracted light with refraction coefficients $t_{ee}$ and $t_{em}$ in the TI take the forms:

\begin{eqnarray}
\label{eq16}
\bm{E} &=& t_{ee}\omega(-k_y\bm{e}_x+k_x\bm{e}_y) + c t_{em}
(p_z\bm{k}-k_2\bm{e}_z), \nonumber\\
\bm{B} &=& t_{ee}(-p_z\bm{k}+k^2\bm{e}_z) +
\frac{t_{em}}{c}\omega(-k_y\bm{e}_x+k_x\bm{e}_y), 
\end{eqnarray}

where $c$ is the relative velocity of light in TI bulk, $\bm{k}=k_x\bm{e}_x+k_y\bm{e}_y$, $k^2=k_x^2+k_y^2$, and $p_z$ is the $z$ component of the wave vector in the TI. From the boundary conditions we deduced the following equations on the $j$th boundary:

\begin{eqnarray}
\label{eq17}
&&1-t^{(j)}_{ee} + r^{(j)}_{ee} = 0, \nonumber\\
&&\frac{p_z}{\sqrt{\varepsilon\mu}}t^{(j)}_{em} + k_z r^{(j)}_{em} = 0, \nonumber\\
&&r^{(j)}_{em}-\sqrt{\frac{\varepsilon}{\mu}}t^{(j)}_{em} =
\left(\phi_j(r^{(j)}_{ee}+1)+2i\Phi_1k_zr^{(j)}_{em}\right), \nonumber\\
&&p_zt^{(j)}_{ee}+{\mu}k_z(r^{(j)}_{ee}-1)=-\mu\phi_jk_zr^{(j)}_{em} \nonumber\\
&&\hspace{0.5cm}+2i\mu(1+r^{(j)}_{ee})(\Phi_1\omega^2-\Phi_2v_F^2k^2).
\end{eqnarray}

For the TM mode injection, one can write similar equations with reflection coefficients $r_{me}$, $r_{mm}$ and refraction coefficients $t_{me}$, $t_{mm}$. Their solutions are given by(we need only the exact form of reflection coefficients):

\begin{eqnarray}
\label{eq18}
&&r_{ee}^{(j)}=-1+\frac{2}{D}{\left(1+\varepsilon\frac{k_z}{p_z}+2\Phi_1 k_z \right)}, \nonumber\\
&&r_{em}^{(j)}=r_{me}^{(j)}=\frac{2\phi_j}{D}, \nonumber\\
&&r_{mm}^{(j)}=1-\frac{2}{D}{\left[\left(1+\frac{p_z}{k_z}\right)+\frac{2}{k_z}\left(\Phi_1\omega_n^2
+\Phi _2v_F^2\bm{k}^2 \right)\right]}, \nonumber
\end{eqnarray}

where the superscript $j$ means the $j$th interface, and

\begin{eqnarray}
\label{eq19}
D&=&\left(1+\varepsilon+\phi^2\right)+\left(\varepsilon
\frac{k_z}{p_z}+\frac{p_z}{k_z}\right)+2\Phi_1\left(k_z+{p_z}\right) \nonumber\\
&&+2\left(\frac{\varepsilon}{p_z}+\frac{1}{k_z}+2\Phi_1\right)\left(\Phi_1\omega_n^2+\Phi_2v_F^2\bm{k}^2\right).
\end{eqnarray}

We note that we have already translated the expressions into Matsubara imaginary time formalism and we assume the influence from permeability can be omitted, $\mu=1$. In imaginary frequency formalism, $k_z=\sqrt{\omega_n^2+k^2}$ and $p_z=\sqrt{\varepsilon\omega_n^2+k^2}$. For practical calculation, we need a form of frequency-dependent dielectric permittivity, which can be modeled by
\cite{Bordagreport2001,Bordagbook}

\begin{equation}
\label{eq20}
\varepsilon=1+\sum_{J=1}^{K}\frac{g_J}{\omega_n^2+\omega_J^2+\gamma_J\omega_n},
\end{equation}

with $K$ oscillators and for each oscillator, the oscillation strength is $g_J$ and oscillation frequency is $\omega_J$, $\gamma_J$ is the corresponding damping parameter. We consider only one oscillator and omit the contribution from damping parameter here, the generalization to multi-oscillator and non-zero damping parameters is straightforward. Then the Casimir energy density at finite temperature can be deduced from Lifshtz formula:

\begin{equation}
\label{eq21}
\frac{E_C}{A} = \frac{1}{\beta}\sum_{n=0}^{\infty}{}'\int
\frac{d^2k}{(2\pi)^2}
\log\det\left(1-\mathcal{R}^{(1)}\mathcal{R}^{(2)}
e^{-2k_z{a}}\right), 
\end{equation}

where the prime in the summation means for the $n=0$ term there contains a prefactor $\frac{1}{2}$, and $\mathcal{R}^{(1,2)}$ are Fresnel coefficient matrices on the surfaces, which take the forms:

\begin{equation}
\label{eq22}
\mathcal{R}^{(j)} = \left(
                \begin{array}{cc}
                  r_{ee}^{(j)} & r_{em}^{(j)} \\
                  r_{me}^{(j)} & r_{mm}^{(j)} \\
                \end{array}
              \right). 
\end{equation}

\begin{figure}
\includegraphics[width=0.45\textwidth]{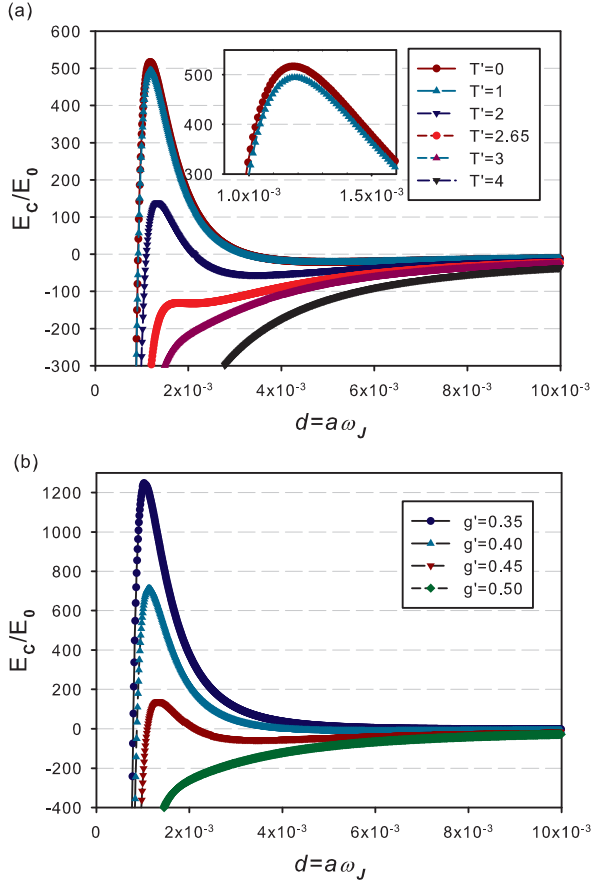}
\caption{(a) Casimir Energy as a function of dimensionless
distance between TIs at different temperatures, here the
dimensionless oscillation strength has been chosen to be
$g'=\sqrt{g_J}/\omega_J=0.45$, surface band gap $|m|=500\omega_J$,
and $T'=T/T_0$, where the definition of $T_0$ has been given in
the context. (b) Casimir Energy as a function of dimensionless
distance between TIs for different oscillation strength
$g'=\sqrt{g_J}/\omega_J$, here temperature $T=2T_0$ and surface
band gap $|m|=500\omega_J$.\label{fig2}} 
\end{figure}

\section{\label{sec4}Results and Discussions}

It is hard to obtain the full analytical expressions of $\Phi_1$, $\Phi_2$ and $\phi$, and a general analyse of the Casimir energy for finite surface band gap at finite temperature seems to be very difficult. Contrast to the usual calculation of Casimir interaction at finite temperature, here the finite temperature correction can be divided into two part, the one part is the difference between integration and discrete summation, the other part is from the finite temperature correction of $\Phi_1$,
$\Phi_2$ and $\phi$, and the widely used Abel-Plana formula
\cite{SaharianarXiv2000,Bordagbook}
does not work here because the integration kernel do have singularities on the right-half complex plane which have been implicit contained in the integral form of $\Phi_1$, $\Phi_2$ and $\phi$.

Here, we are only concerned with the critical surface band gap for repulsive Casimir interaction. As shown in Ref.
\cite{LChenprb2011},
the critical surface band gap is much greater than room temperature, $m_c\gg300K$, so low temperature expansion is a good approximation. In practical calculation, we
sum over the first several terms of Eq.\ref{eq21} and use the integration over the rest regime to approximate the summation with corrections evaluated by Euler-Maclaurin formula
\cite{Bordagbook}.

Before the detailed discussion of results obtained, we make a note on units chosen in this paper, we have set the Plank constant and velocity of light in vacuum to 1, and we choose oscillation frequency $\omega_J$ as the unit of energy. The unit of Casimir energy and temperature are $E_0=\omega_J^3/8\pi^2$ and $T_0=\omega_J/2\pi$ respectively.

\begin{figure}
\includegraphics[width=0.45\textwidth]{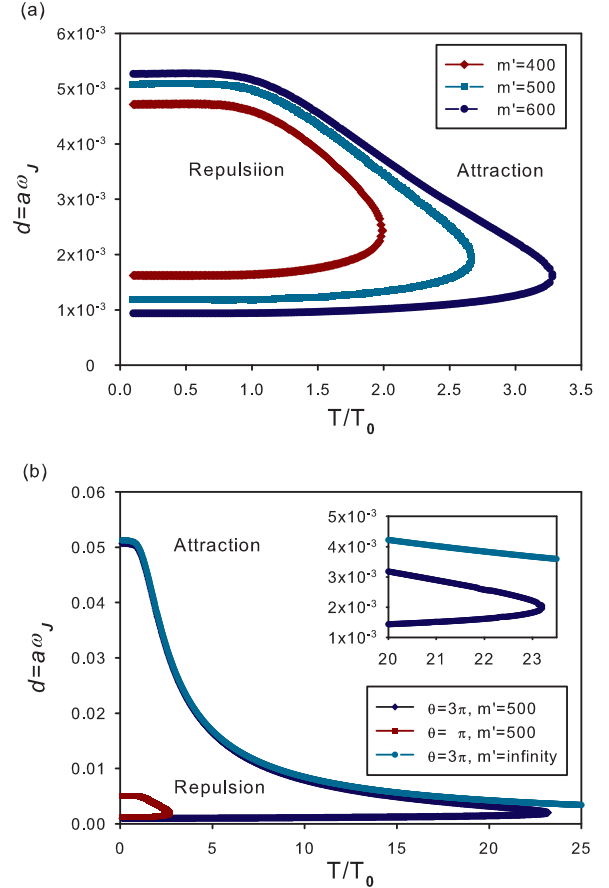}%
\caption{Equilibrium distance of Casimir force as a function of
temperature for (a) different surface band gap and (b) different
topological magnetoelectric polarizabilities. The dimensionless
oscillation strength has been chosen to be
$\sqrt{g_J}/\omega_J=0.45$, definition of dimensionless surface
band gap and temperature are given by $m'=|m|/\omega_J$ and
$T_0=\omega_J/2\pi$ respectively. Insert shows the detailed
difference between $m'=500$ and infinity surface band gap.\label{fig3}}
\end{figure}

\begin{figure}
\includegraphics[width=0.45\textwidth]{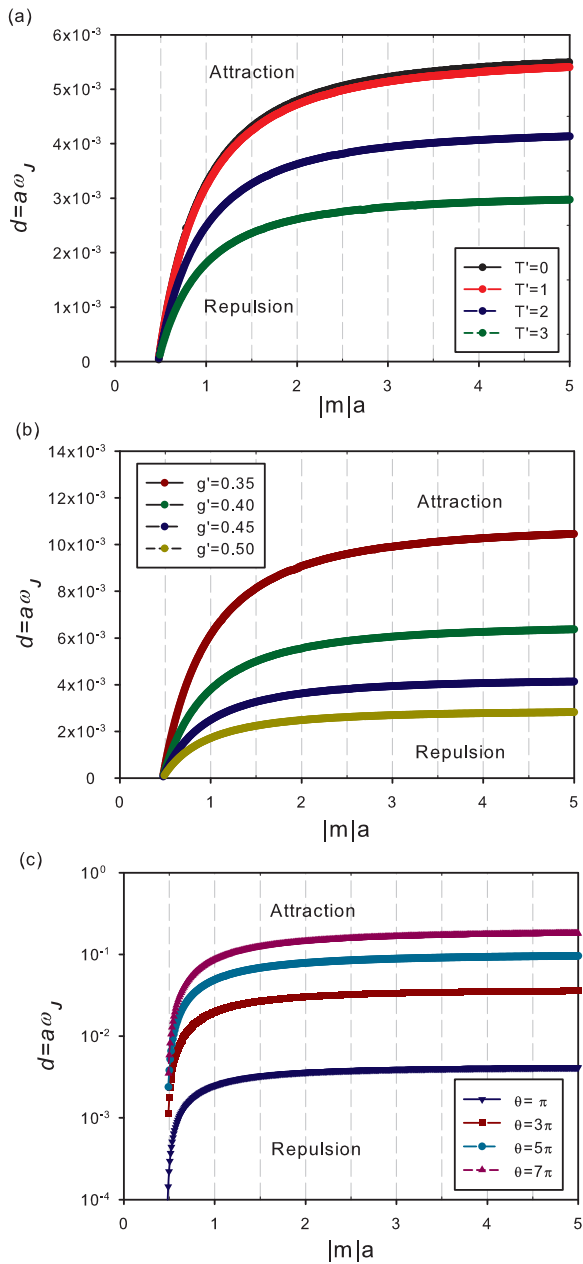}
\caption{Boundary of repulsive and attractive Casimir interaction
as a function of dimensionless distance $d=a\omega_J$ and product
$|m|a$ for (a) different temperature $T/T_0$, (b) different
oscillation strength $g'=\sqrt{g}/\omega_J$ and (c) different
topological magnetoelectric polarizabilities
$\theta$.\label{fig4}}
\end{figure}

First, we obtain Casimir energy as a function of distance between TIs for different temperature, in Fig.\ref{fig2}(a), and different oscillation strength, in Fig.\ref{fig2}(b). Fig.\ref{fig2}(a) is one of the major results, which shows that increasing temperature will depress the repulsive peak and reduce the distance between
local maximum and minimum points of Casimir energy, and at the critical temperature $T_C$, they equal to each other and the repulsive Casimir interaction vanishes, for given parameters $\sqrt{g_J}/\omega_J=0.45$ and $|m|=500\omega_J$, the critical temperature $T_C\sim2.65T_0$, as shown by the red circle dotted line. Such a result is well understood because in the high temperature limit, Casimir interaction will tend to the classical limit and the majority contribution is the zero-frequency term and
the quantum fluctuation from surface Dirac fermions is suppressed. Repulsive Casimir interaction is also quantitatively influenced by oscillation strength of electromagnetic wave in TIs bulk, small oscillation strength will decrease the attractive Casimir force from TI bulk and profitable for repulsive peak, as shown in
Fig.\ref{fig2}(b).

Second we give the local maximum and minimum points of Casimir energy(they are both equilibrium distances of Casimir force) as a function of temperature for different surface band gap, in Fig.\ref{fig3}(a), which shows that larger surface band gap will make the critical temperature higher and repulsive distance larger. However, such a exertion seems to be difficult to achieve and produce little effect compared with increasing topological magnetoelectric polarizability. We also give the local maximum and minimum points of Casimir energy for topological magnetoelectric polarizability $\theta=3\pi$ by introducing multi-fermions on TI surfaces, in Fig.\ref{fig3}(b), which shows that large topological magnetoelectric polarizability will remarkably increase the scope of repulsive Casimir force.

Then, as a competition, we also give the equilibrium distance as a function of temperature for infinite surface band gap limit $|m|\rightarrow\infty$, as in Fig.\ref{fig3}(b), which shows that, at low temperature, the larger equilibrium distance of Casimir interaction for a large but finite surface band gap $|m|=500\omega_J$ is very close to the equilibrium distance of Casimir interaction for infinite surface band gap, however, at the critical temperature, $T_C\sim 23.5T_0$, the larger equilibrium distance and smaller equilibrium distance equal, so Casimir force will always attractive when $T>T_C$.

Finally, similar to the zero temperature case, we also give the boundary of attractive Casimir interaction and repulsive Casimir interaction as a function of dimensionless distance $d=a\omega_J$ and product $|m|a$ for different temperature, different oscillation strength and different topological magnetoelectric polarizabilities, in Fig.\ref{fig4}, and find that the critical product $m_c\,a\sim1/2$ is still valid, which show that in the short distance limit Casimir interaction is dominated by surface fermions, where the topological response of surface fermions gives a repulsive Casimir interaction and the electromagnetic dynamical response of surface fermions give an attractive Casimir interaction, these contributions have the same magnitude.\\

\section{\label{sec5}Conclusions}

In this paper, we calculate the Casimir interaction between TIs with opposite topological magnetoelectric polarizability and finite surface band gap at finite temperature, and find that, finite temperature will quantitatively affect Casimir interaction, if Casimir interaction is repulsive for proper distance at zero
temperature, rising temperature will depress the repulsive peak and at a critical temperature, the Casimir interaction will be attractive for any distance between TIs. We also find that the estimation relationship $m_c{a}\sim{1/2}$ for critical repulsive Casimir interaction is valid for different temperature, different oscillation strength and different topological magnetoelectric polarizabilities, which is useful for practical research of repulsive Casimir interaction between TIs.

\begin{acknowledgments}
We acknowledge helpful discussions on program with Xiaosen Yang and Mengsu Chen. This work is supported by NSFC Grant No.10675108.
\end{acknowledgments}

\bibliographystyle{apsrev4-1}
\bibliography{reference}

\end{document}